\documentclass[journal ]{new-aiaa}
\usepackage{geometry}
\usepackage[utf8]{inputenc}
\usepackage{textcomp}
\usepackage{mwe}
\usepackage{graphicx}
\usepackage{amsmath}
\usepackage[version=4]{mhchem}
\usepackage{siunitx}
\usepackage{longtable,tabularx}
\usepackage{xcolor}
\usepackage{color}
\usepackage{url}

\usepackage{todonotes}
\usepackage{changebar}
\usepackage{lipsum}
\usepackage{ulem}
\usepackage{soul,xcolor}
\setstcolor{blue}

\usepackage{caption}
\usepackage{subcaption}

\title{Investigation of Aerodynamic Characteristics of a Generic Transport Aircraft in Ground Effect Using URANS Simulations}

\author{Mohamed Sereez,\footnote{Lecturer,  School of Future Transport Engineering, MRAeS}  Nikolay B. Abramov,\footnote{Senior Lecturer, School of Engineering and Sustainable Development, MAIAA} and 
Mikhail G. Goman\footnote{Professor of Dynamics, School of Engineering and Sustainable Development, FRAeS, SMAIAA}}
\affil[1]{Coventry University, Coventry, CV1 5FB }
\affil[2,3]{De Montfort University, Leicester, LE1 9BH, United Kingdom}

\begin{document}

\maketitle

\begin{abstract}

This paper focuses on computational prediction of  aerodynamic and the flow field characteristics for NASA Common Research Model (CRM) in it's High-Lift (HL) configuration in close proximity to the ground. The URANS simulation with the Spalart-Allmaras (SA) turbulence model is checked for the quality of the generated mesh and compared with the available wind tunnel data. The obtained simulation results in the immediate vicinity of the ground demonstrate significant changes in the longitudinal and lateral-directional aerodynamic characteristics in aircraft banked positions, which is important for a better understanding of aircraft landing in crosswind conditions.

\end{abstract}

\section{Introduction}

Close-to-ground aerodynamics are important for development of advanced flight dynamics models with six degrees of freedom during takeoff and landing in crosswind conditions.
The risk of Runway Excursion (RE) and Approach and Landing Accident Reduction (ALAR) can be further reduced or mitigated by incorporating the data acquired from computational and experimental investigations of ground effect aerodynamics into the improved flight dynamics models. 
This objective is now classified as an important research area in the Future Sky program of the Association of Establishments of European Research in Aeronautics, EREA.

According to worldwide accident statistics for the commercial aircraft fleet, fatal accidents for the period 2006-2015 due to abnormal runway contact (ARC) and runway excursion (RE) rank second after the Loss Of Control In-Flight (LOC-I) \cite{Boeing}.
To reduce such risks, Approach and Landing Accident Reduction  (ALAR) is one of the main goals of the Flight Safety Foundation (FSF) \cite{FSF}.
In order to achieve this goal, pilots need to be trained on flight simulators equipped with improved six-DOF flight dynamics models representing the accurate transformation of flight dynamics during approach and landing phases of flight.

Computational simulations of ground effect aerodynamics using the URANS equations is now becoming an important addition to the wind tunnel experimental methods. Simulation results based on the URANS equations are generally in good agreement with wind tunnel data in ground effect for symmetric longitudinal attitudes, and this motivates us to extend the simulation to more general aircraft attitudes. To better understand the lateral-directional dynamics during landing, this paper focuses on URANS simulations of an aircraft's aerodynamics near the ground for its attitude with non-zero bank angles.

In symmetrical positions, In-Ground-Effect (IGE) aerodynamics is characterized by an increase in the lift coefficient $C_L$, positive increments in the pitching moment coefficient $C_m$ acting in the "nose-up" direction, and a significant decrease in the drag coefficient $C_D$. The increase in $C_L$ is mainly due to the chord-dominated ground effect, and the increase in pitching moment $C_m$ is due to a significant increase in pressure on the lower surface of the fuselage in close vicinity to the nose \cite{DLRF6Geff,MSereezImpactofGround}.
The other studies of single-element airfoils and wings are quite well-represented \cite{DOIGAIAAGround2,MahongAIAAGRound3,N4412GEff}, also showing similar results in ground effect aerodynamics.

The aerodynamics of multi-element airfoils in close proximity to the ground shows a significantly different aerodynamic behavior \cite{Partha30P30N,QinMultiCFDGround,OJFD_GroundEff,DatCom,HoakUSFGround}. 
Instead of an increase in the lift coefficient in close proximity to the ground, as observed for single-element airfoils,  the multi-element airfoil "30P30N", for example, shows a noticeable drop in the lift coefficient 
\cite{Partha30P30N,OJFD_GroundEff}. 
 Identical trends of large loss of the lift coefficient in-ground-effect are observed for the multi-element "L1T2" airfoil configuration \cite{QinMultiCFDGround}. It is shown that with increased proximity to the ground,  the losses of pressure on the suction side of the airfoil are more than the increase of pressure on the lower side, and the separation zone of the upper flap surface becomes bigger. In other words, the "positive ground effect" dominated by the increased pressure on the lower surface still exists to some extent but is reversed by the large losses of pressure on the upper side \cite{HoakUSFGround,DatCom}.

\begin{figure}[htb!]
\centering
\includegraphics[width=0.7\textwidth]{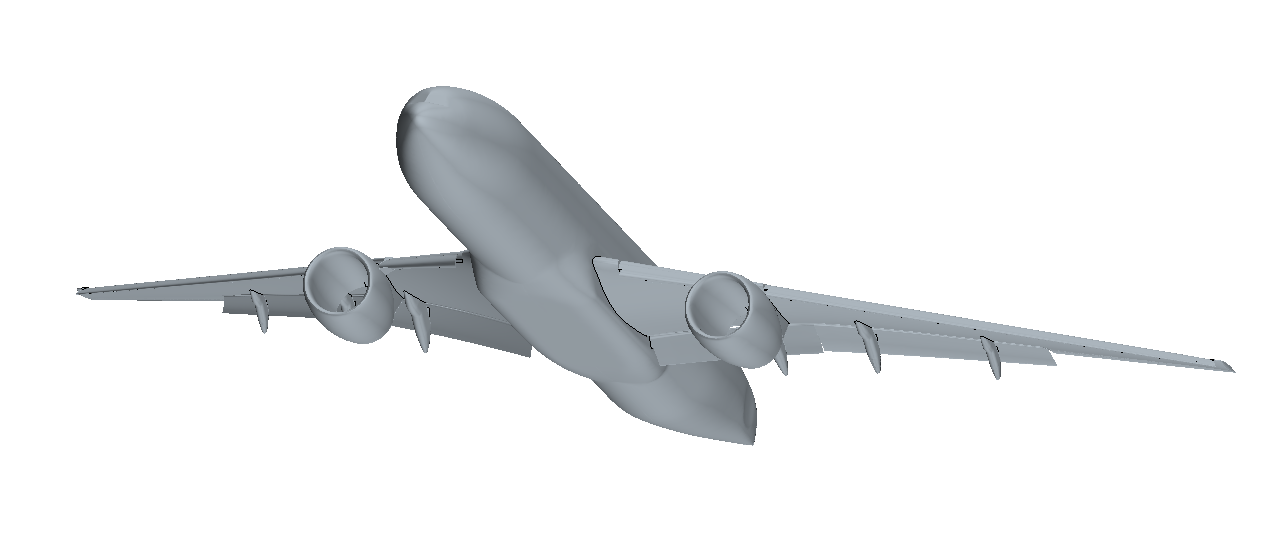}
\caption{Geometry of the CRM High-Lift nominal configuration.}
\label{Fig1-Geometry}
\end{figure}

This paper focuses on the investigation of aerodynamics of a generic transport airliner, the NASA Common Research Model (CRM) in its Wing-Body High-Lift (WB-HL) configuration in close-proximity to the ground. The Reynolds number and Mach number are taken as $Re=5.49\times10^6$ and $M=0.2$ and the height to chord ratios, characterizing the closeness to the ground, as $h/c=1.5,1.35$ and $1.0$. The choice of flight speed is justified as it represents the landing configuration speed of about $130-135$ knots \cite{FSF}. To allow comparison of our simulations with wind tunnel data and published CFD results, the Reynolds number is equal to $Re=5.49\times10^6$ as used in the high-lift prediction workshop \cite{HLPW}. The geometry of the CRM-WB-HL configuration and the computational grids generated are shown in  Figures \ref{Fig1-Geometry} - \ref{Fig3-Grid}. 
Although the grids for the CRM-WB-HL model are available from the high-lift prediction workshop \cite{HLPW}, following the guidelines provided in \cite{HLPW} our own computational grids were generated in order to  facilitate "overset/chimera" simulation methods. 

The paper is organized as follows. The computational framework including the model geometry, governing equations, boundary conditions, grid generation, and numerical solver settings are presented in the computational framework in Section \ref{2-CFDsetup}. The computational validation of the simulation results is conducted against experimental results in Out-of-Ground Effect (OGE) condition and mesh independence studies for both the OGE and In-Ground-Effect (IGE)  are shown in Section \ref{3-CFDValidation}. The obtained simulation results along with relevant discussions and post-processing are presented in Section \ref{4-CFDresults} and are followed by the concluding remarks in Section \ref{5-CFDConclusions}.

\section{Computational Framework}
\label{2-CFDsetup}
This section presents the adopted computational framework, which includes the geometry of the CRM-WB-HL configuration, the methodology adopted for meshing, and the general setup for ground-effect case studies.

\subsection{Geometry and setup for ground effect}
The geometry for the CRM-WB-HL configuration is an extension of the CRM wing-body geometry \cite{UnifiedCRM} used in the drag prediction workshop \cite{DragPW4}. The CRM wing is made up of a thin super-critical airfoil with an aspect ratio $AR\mbox{=}9$ and a taper ratio of $0.25$.
The high-lift configuration of CRM\cite{CRMHL} which is used in this paper for evaluation of ground effect is the nominal configuration provided by the high-lift prediction workshop committee in \cite{HLPW}. 
The high-lift CRM wing has the same reference geometry parameters as the CRM wing-body configuration, mean aerodynamic chord $c_{ref}=7.0m$, wing span, $b=58.76m$, and reference area $S_{ref}=383.65m^2$. 
The inboard and outboard flaps are deflected with angles of $40^\circ$ and $37^\circ$ respectively and the slat is deployed at an angle of $30^\circ$. Table. \ref{Tab:Table1} includes all reference parameters for the CRM-WB-HL configuration. The original CAD model and additional information on the geometry are available in \cite{HLPW,CRMHL,CRMHLexp}.

\begin{table}[htb!]
\begin{center}
\captionof{table}{Reference data for the CRM-WB-HL (full model)  \label{Tab:Table1}}
\begin{tabular}{cc}
\hline
Wing span, $b$ & $58.76\, m$ \\
Mean aerodynamic chord $MAC$: $c_{ref}$ & $7\, m$ \\
Inboard/Outboard flap angles & $40^\circ/37^\circ$ \\
Slat angle & $30^\circ$ \\
Reference area  $S_{ref}$ & $383.68\,m^2$ \\
Wing aspect ratio, $AR$ & $9.0$ \\
Moment reference point $C_g$ & $X\mbox{=}33.67\, m, Y\mbox{=}0, Z\mbox{=}4.52\, m $ \\
\hline
\end{tabular}
\end{center}
\end{table}

\begin{figure}[htb!]
\centering
\includegraphics[width=1.0\textwidth]{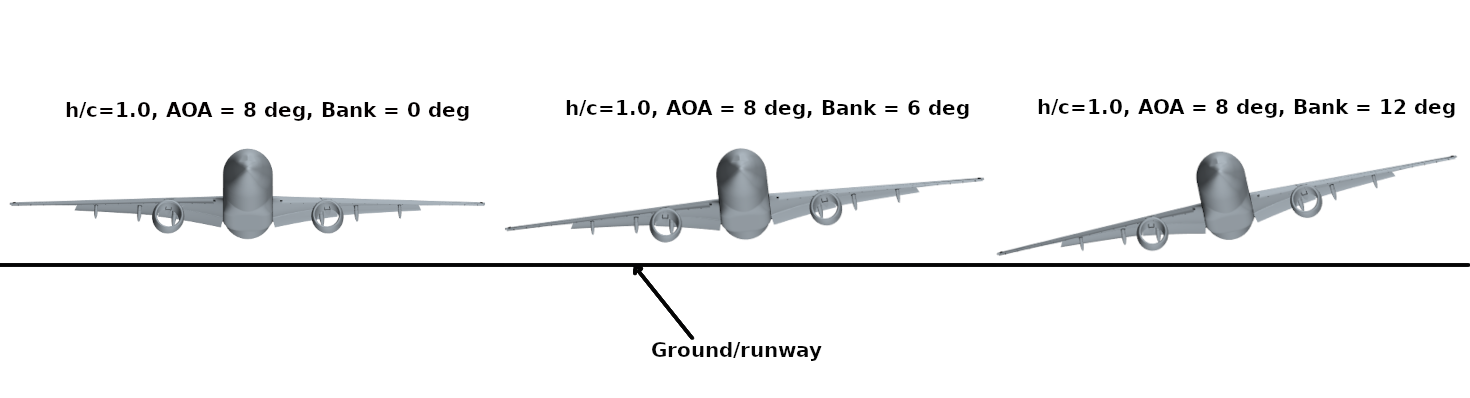}
\caption{Considered aircraft positions for investigation of the lateral-directional aerodynamics in close proximity to the ground at $h/c=1.0$ and $\alpha=8.0^\circ$.}
\label{Fig2-GeometryBank}
\end{figure}

The computational domain is a virtual wind tunnel type rectangular shape box with a scale factor of $1000 c_{ref}$ away from the aircraft model in every direction, except for the cases of ground effect where the negative $Z$ direction had to be fixed in order to set up the correct non-dimensional ratio $h/c$ defining the closeness to the ground.
The usual boundary conditions used in incompressible flow simulations i.e. the velocity inlet with fixed value boundary condition, and pressure outlet with zero gradient condition at the outlet were employed.
For the wind tunnel walls, a slip boundary condition is employed while a no-slip boundary condition was used for the aircraft surfaces.
In the case of close proximity to the ground, a prism boundary layer was used and a "moving wall" boundary condition was allocated with specified relative velocity in order to accurately capture the ground effect.
In addition, for investigation of the ground effect, the desired flight attitude is obtained by rotation around the moment reference point given in Table \ref{Tab:Table1}.
An example of a setup for studying ground effect aerodynamics at $h/c=1.0$, $\alpha=8.0^\circ$ and various roll angles $\phi$ is shown in Fig. \ref{Fig2-GeometryBank}.

\subsection{Governing equations}
The Navier--Stokes (NS) equations governing incompressible fluid flow are 
the continuity equation 
\begin{equation}
\label{Eq1-Cont}
\nabla \cdot\mathbf{U} = 0
\end{equation}
and the momentum equation
\begin{equation}
\label{Eq2-Momentum}
\frac{\partial \mathbf{U}}{\partial t} + (\mathbf{U}\cdot\nabla)\mathbf{U}  - \nu \nabla^2 \mathbf{U} = -\frac{\nabla \mathbf{p}} {\rho}
\end{equation}		

The computational resources required for Direct Numerical Simulations (DNS) of Eqs.~(\ref{Eq1-Cont}) and (\ref{Eq2-Momentum}), especially for flow conditions with high Reynolds numbers, usually exceed currently available computational capabilities.
Instead the Unsteady Reynolds-Averaged-Navier-Stokes (URANS) equations are solved, in which the Reynolds stresses arising as a result of averaging the fluctuating velocities are described by some additional empirical equations either algebraic or differential to represent an appropriate turbulence model. Most turbulence models for the URANS equations are based on the concept of eddy viscosity, which is equivalent to the kinematic viscosity of a fluid, to describe turbulent mixing or flow momentum diffusion ~ \cite {OpenFOAM_4_CFD}. 
Reynolds stresses arising in the URANS equations due to time averaging are described in linear turbulence models under the Boussinesq assumption:
\begin{equation}
\label{Eq3-Boussinesq}
\tau_{ij} = 2 \nu_t \!\left(S_{ij} -  \frac{1}{3}\frac {\partial u_k}{\partial x_k} \delta_{ij}\right) - \frac{2}{3}\rho k \delta_{ij}1
\end{equation}

\subsection{Grid generation}
For the purposes of this paper, overset type computational grids are generated using Siemens StarCCM+ software which is well known for its robust overset/chimera methods and simulation capabilities \cite{ccmS}. The reason for generating our own grids is twofold. First, it allows us to modify flight attitudes by manipulating the overset mesh enclosing the aircraft in close proximity to the ground. This is necessary as changing the wind velocity vector is not a viable option for the ground effect simulations. Secondly, to simulate the dynamic mesh for aircraft oscillatory motions for computation of unsteady aerodynamic derivatives, which is planned for a later stage of this study. 

The overset mesh region is a close-bound box around the CRM-WB-HL model $2.0\, m$ away in every direction, enabling it to go for reduced proximity to the ground characterized by the non-dimensional ratio $h/c$. The background mesh is the grid generated for the wind tunnel. Further refinements are placed in the wake of the aircraft in both the background mesh and the overset mesh zone. Although not strictly required, conformal mesh sizes are used on the overlapping region between the overset box surface and the background mesh interface of both regions. This enables smooth interpolation of conserved flow variables between the two mesh zones. The cell area and volume in the immediate vicinity of this overlapping region or the "overset interface" region vary by $1\mbox{-}2\%$ between the two zones i.e. the overset zone and the wind tunnel mesh. A coarse grid is initially generated with $8$ million elements. This grid is then uniformly scaled to produce further refined grids. The resulting computational grids are in the range of  $8$-$38$ million elements.

\begin{figure}[htb!]
\centering
\fbox{\includegraphics[width=0.7\textwidth]{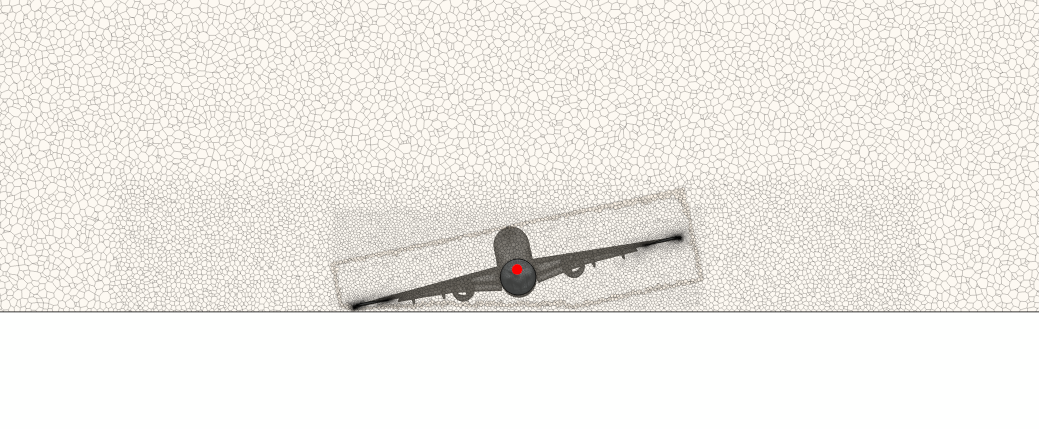}}
\fbox{\includegraphics[width=0.7\textwidth]{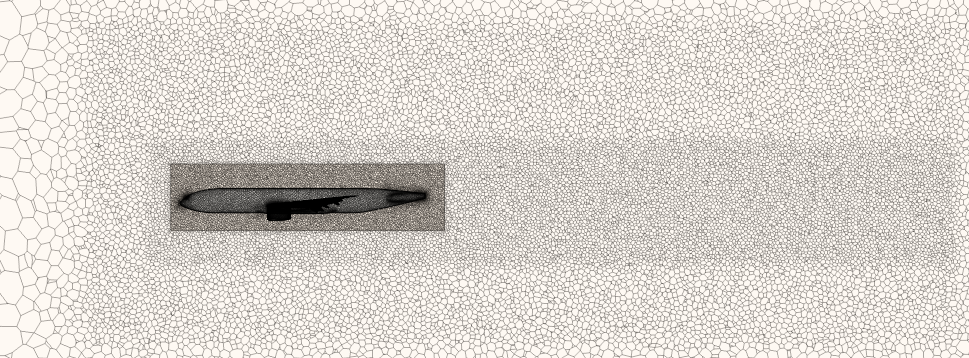}}
\caption{Grid generated for the NASA CRM High-Lift Nominal configuration.}
\label{Fig3-Grid}
\end{figure}
After a fully-fledged grid independence study a mesh size of $16$ million elements was chosen for the initial case studies which consist of a) the validation case study against the experimental results and b) the study of ground effect with varying $h/c$ ratios. A cut-plane view of the grid resolution in the front and the side view is shown in Fig. \ref{Fig3-Grid}. Before moving on to the final test matrix involving the study of ground effect simulations with considered bank angles, another grid independence study was conducted with a mesh size of $12$ million elements. The obtained results proved that a $12$ million element mesh is sufficient to resolve the aerodynamic characteristics at angles of attack below $10$ degrees. A detailed description of the above-mentioned grid independence study results will be presented and discussed further in the validation section of this paper. After careful consideration, the grid size of $12$ million was chosen to continue the investigation of the impact of the ground effect on the lateral-directional aerodynamics.

\subsection{Numerical solver settings}
The numerical solver settings used in this study are consistent with changes brought to the settings depending on the type of simulation. For low angles of attack with $\alpha <10^\circ$, the steady formulation with the Reynolds Averaging of Navier Stokes Equations (RANS) was employed. The RANS formulation was used along with the well-known Spalart-Allamaras \cite{SA} turbulence model. At moderate to high angles of attack $\alpha>10^\circ$ the type of simulation was switched to the Unsteady Reynolds Averaged Navier-Stokes (URANS) formulation as it allows more accurate simulation of separation zones and vortex formations.
The continuity, momentum and turbulence equations were solved in a segregated manner using the Algebraic Multi-Grid (AMG) solver with an inner tolerance of $0.001$. This implies that the solver attempts to solve the Matrix holding the discretized finite volume coefficients to an order of 3 decades. The V and F cycle approach was employed with $2$ pre and post-sweeps in order to maximize convergence. For the multigrid method, the restriction tolerance was kept at $0.9$ and the prolongation tolerance was kept at $0.5$. It was also necessary to employ under-relaxation of conserved variables, especially in ground effect simulations due to the highly unsteady nature of the flow field.

Even though the grids were specially formulated with the low wall $Y+$ approach i.e. $Y+<1.0$, Star-CCM+ recommends the use of "All $Y+$ treatment". With this kind of wall treatment, the boundary layers where the low wall $Y+$ criterion was satisfied are resolved with the low Y+ approach and the boundary layers which violates the low $Y+$ criterion are resolved with the wall treatment method. The gradients of the conservative flow variables are solved with a second-order accuracy with the use of the Venkatakrishnan limiter \cite{Venkata} omitting spurious oscillations in the flow field. The SA turbulence model was used with the "Curvature Correction" option based on presentations made in \cite{HLPW} where it was shown in that this approach gives a good match against experimental results.

\section{Validation of the Computational Framework}
\label{3-CFDValidation}
The computational framework described above was validated in two stages. The first stage involved validation against the experimental results using the medium grid of size $16\times10^6$ elements. Even though the interest was in the low-angle of attack region i.e. $\alpha<10^\circ$, the obtained computational results were validated against the wind tunnel data up until the stall angle of $\alpha_{s}=22.0^\circ$. In the second stage, due to the large test matrix employed in ground effect simulations, a coarse-medium grid of size $12\times10^6$ elements was used to rerun the simulations that were initially done with the medium grid. The obtained results for the coarse-medium grid and the medium grid are generally in good agreement with each other in both OGE and IGE simulations.

\subsection{Validation against experimental results using medium grid}
\begin{figure}[htb!]
\centering
\includegraphics[width=0.52\pdfpagewidth]{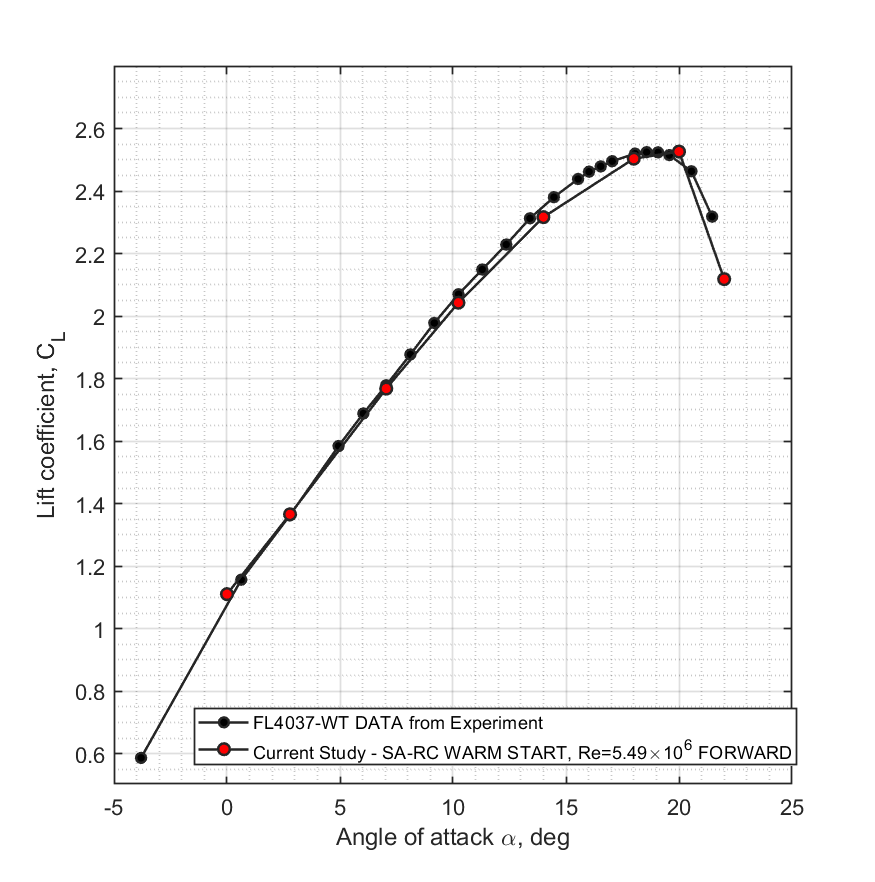}
\includegraphics[width=0.49\textwidth]{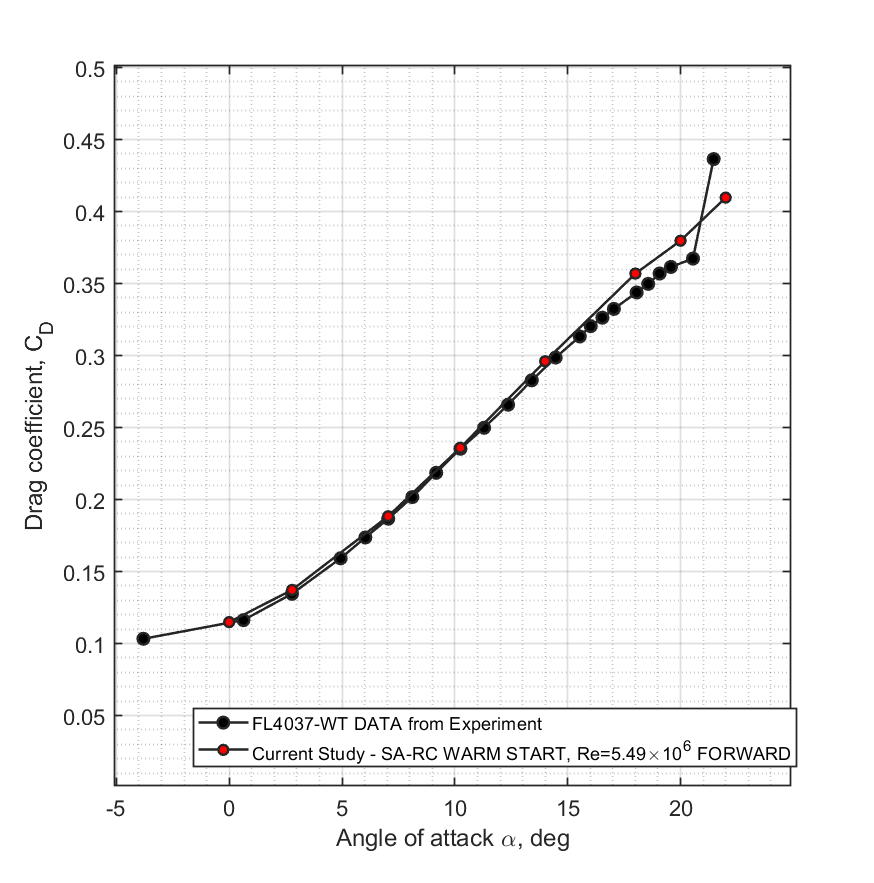}
\includegraphics[width=0.49\textwidth]{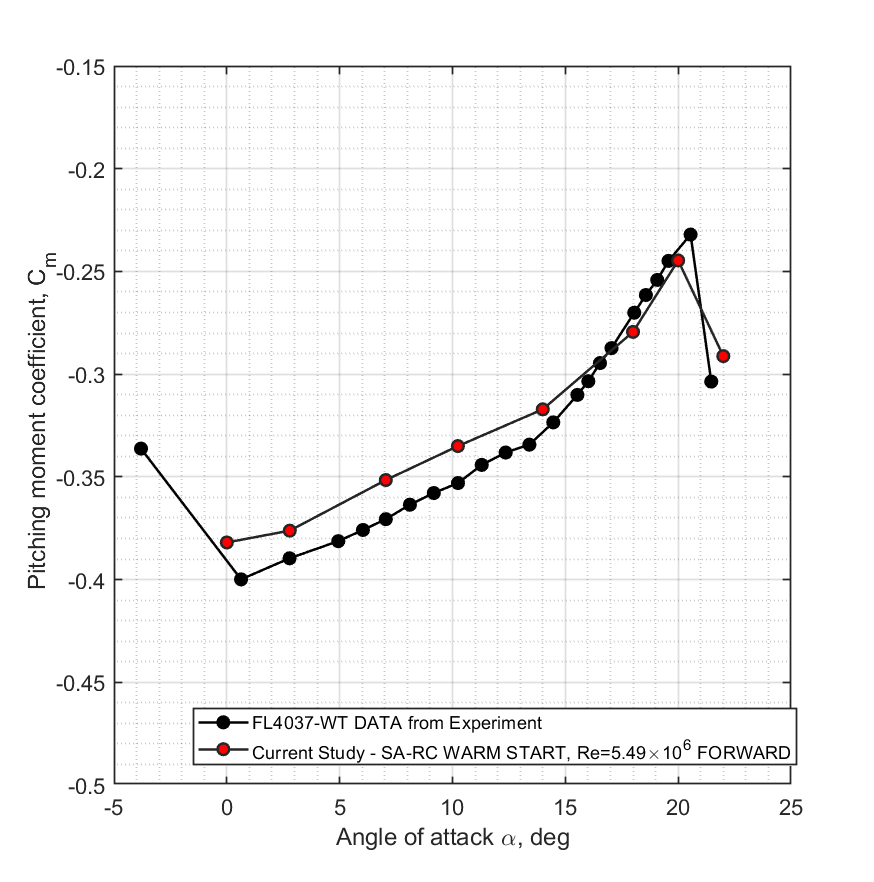}
\caption{Validation of CFD simulation results against the wind tunnel data for Out-of-Ground-Effect (OGE) at $Re=5.49\times10^6$ and $M=0.2$  note: experimental results are from \cite{HLPW}.}
\label{Fig4-ExpValidation}
\end{figure}

The computational framework with the Spalart-Allmaras (SA) turbulence model \cite{SA} for the CRM-WB-HL model was successfully validated against available wind tunnel data (no ground effect) in free stream flight conditions at Reynolds number of $Re=5.49\times10^6$ and Mach number $M=0.2$ (Figure \ref{Fig4-ExpValidation}).  For this study, the medium grid with a total of $16$ million elements was used, and the simulations were run on the Zeus HPC cluster facility at Coventry University \cite{ZeusHPC}.

The obtained computational results for the lift coefficient, drag coefficient, and pitching moment coefficient are shown against the experimental data from the QinetiQ Five-Meter Pressurized Low-Speed Wind Tunnel FL4037 in Fig. \ref{Fig4-ExpValidation}. These wind tunnel experimental data are available to download from the high-lift prediction workshop site \cite{HLPW}.
The information on the wind tunnel setup and experimental results for the high-lift configuration can be found in \cite{CRMHLexp}. 
Fig. \ref{Fig4-ExpValidation} shows that the obtained computational simulation results obtained for the lift coefficient $C_L$ and drag coefficient $C_D$ are in very good agreement with the experimental data at low angles of attack and in the stall region predicting quite accurately the maximum lift coefficient and the drop in the lift coefficient after the stall \cite{HLPW,CRMHLexp}. There is a slight shift for the pitching moment coefficient $C_m$ at low angles of attack, however, the general trend in variation of the pitching moment is captured accurately and the stall angle matches with wind tunnel data.

\begin{figure}[htb!]
\centering
\includegraphics[width=0.8\pdfpagewidth]{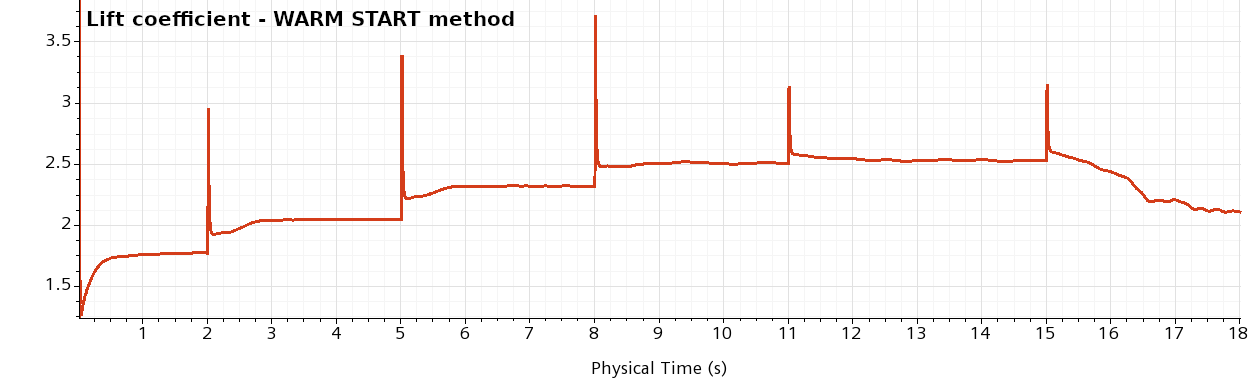}
\caption{Convergence of lift coefficient $C_L$ using the WARM START method for Out-of-Ground-Effect (OGE) at $Re=5.49\times10^6$ and $M=0.2$.}
\label{Fig5-ValidationB}
\end{figure}

The "WARM START" simulation method in Fig. \ref{Fig4-ExpValidation} refers to a technique in aerodynamic computational simulations where the converged flow field of the previous angle of attack's data is used as the initial solution for the flow field for simulation at a new angle of attack. Rather than starting from a non-converged field, the "WARM START"  method allows capturing the forward loop quite well. This technique is good for capturing the static hysteresis phenomena as the backward loop can be simulated robustly as well. The convergence process of the lift coefficient $C_L$ in the Out-of-Ground-Effect (OGE) simulations using the "WARM START" method is shown in Fig.\ref{Fig5-ValidationB}.

\subsection{Validation of mesh resolution in and out of ground effect}
To save computational resources while maintaining the desired accuracy, a coarse-medium grid of size $12$ million elements is generated and compared with the simulation data obtained at low angles of attack using the medium grid containing $16$ million elements. The comparisons are carried out for both In-Ground-Effect (IGE) and Out-of-Ground-Effect (OGE).

\begin{figure}[t!]
\centering
\includegraphics[width=0.9\textwidth]{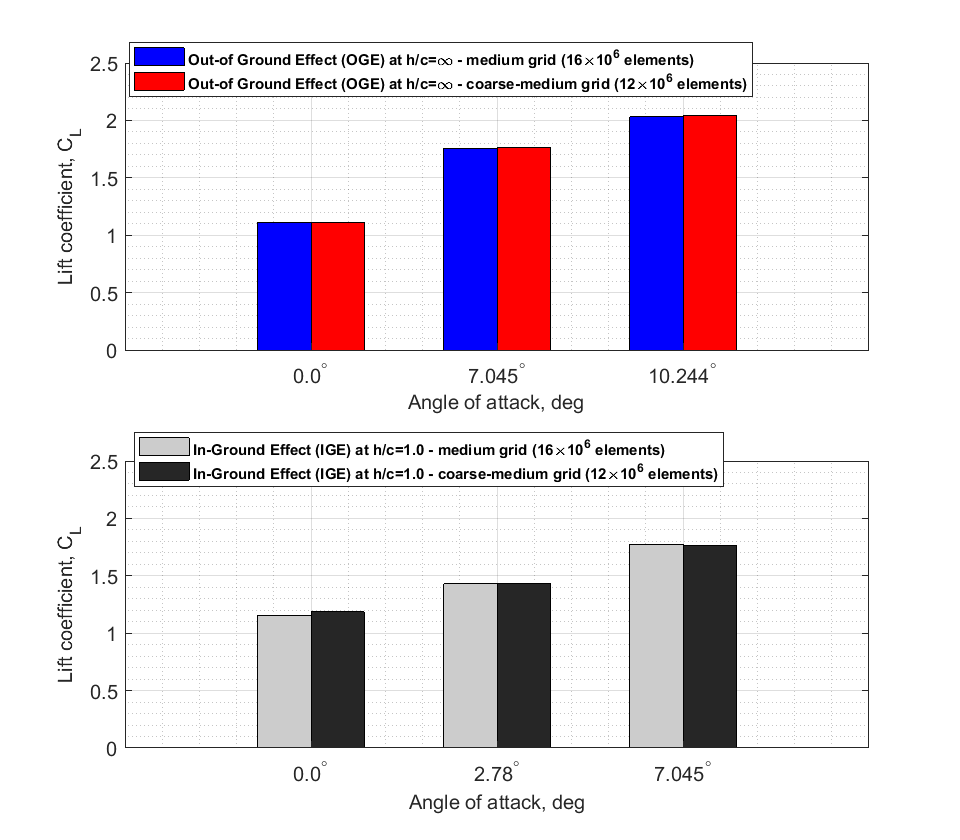}
\caption{Mesh Independence study for the OGE and IGE simulations at $Re=5.49\times10^6$ and $M=0.2$.}
\label{Fig6-MeshIndependence}
\end{figure}

The obtained computational results using the Star-CCM+ CFD package are shown in Fig. \ref{Fig6-MeshIndependence} in the form of bar charts. This method displays the differences between the convergence results for both the OGE and IGE cases for the two grids as the most informative. 
For the OGE cases, the chosen angles of attack are below $\alpha=10.244^\circ$. This is because, for the ground effect simulations, the focus is placed on the trim angle of attack $\alpha_{trim}$ for landing configuration, which is between $0-10^\circ$ for a generic transport airliner (see in \cite{HeffleyHandlingQualities}, p.216). The lift coefficient $C_L$ obtained for the coarse-medium and the medium grid are in perfect agreement with each other. For the IGE simulations with $h/c=1.0$ the lift coefficient $C_L$ obtained at $\alpha=0, 2.78^\circ$ and $10^\circ$ are in good agreement as well.

\section{Simulation results and discussion}
\label{4-CFDresults}

The computational results for aerodynamic characteristics and flow field parameters obtained for the CRM-WB-HL configuration at $Re=5.49\times10^6$, $M=0.2$, and different height-to-chord ratios $h/c$ are discussed in this section. The effect of ground on the longitudinal aerodynamic characteristics at symmetric attitudes with zero bank angles is presented first and followed by the ground effect on the lateral-directional aerodynamics at non-zero bank angles. A constant non-dimensional height to chord ratio of $h/c=1.0$ is taken as a case of close proximity to the ground, and the angle attack of $\alpha=8.0$  is combined with various bank angles $\phi=0, 4, 6, 8, 10^\circ$. Sample attitudes of the aircraft in the ground proximity are shown in the section Computational Framework in Fig. \ref{Fig2-GeometryBank}.

\subsection{Ground effect on the longitudinal aerodynamics}

\begin{figure}[h!]
\centering
\includegraphics[width=0.7\textwidth]{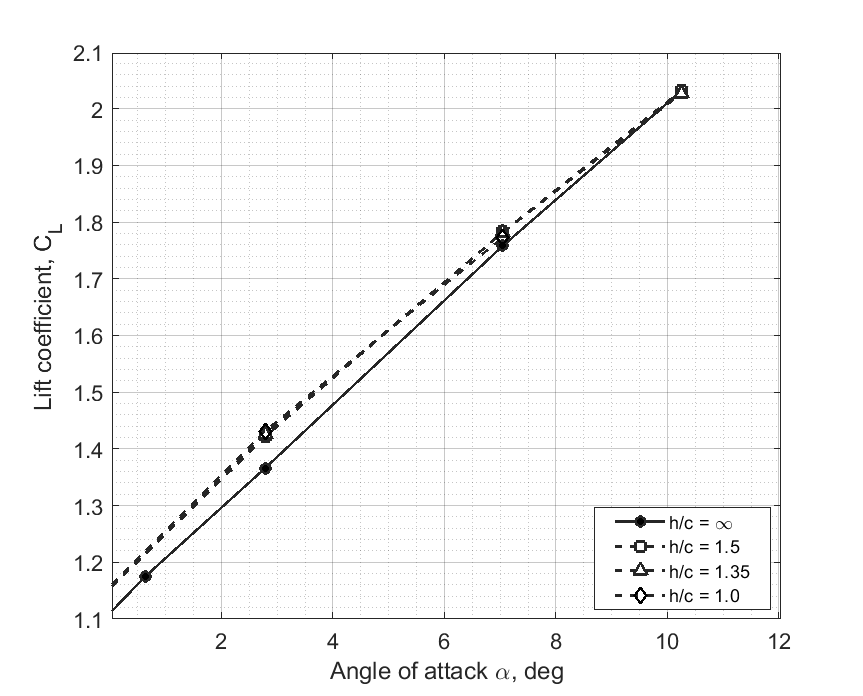}
\includegraphics[width=0.46\textwidth]{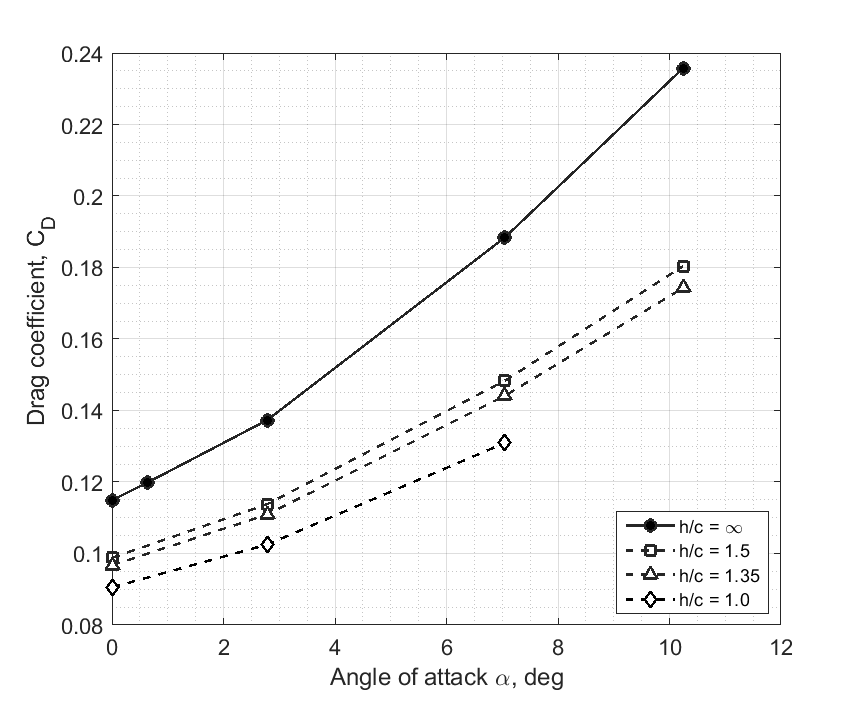}
\includegraphics[width=0.46\textwidth]{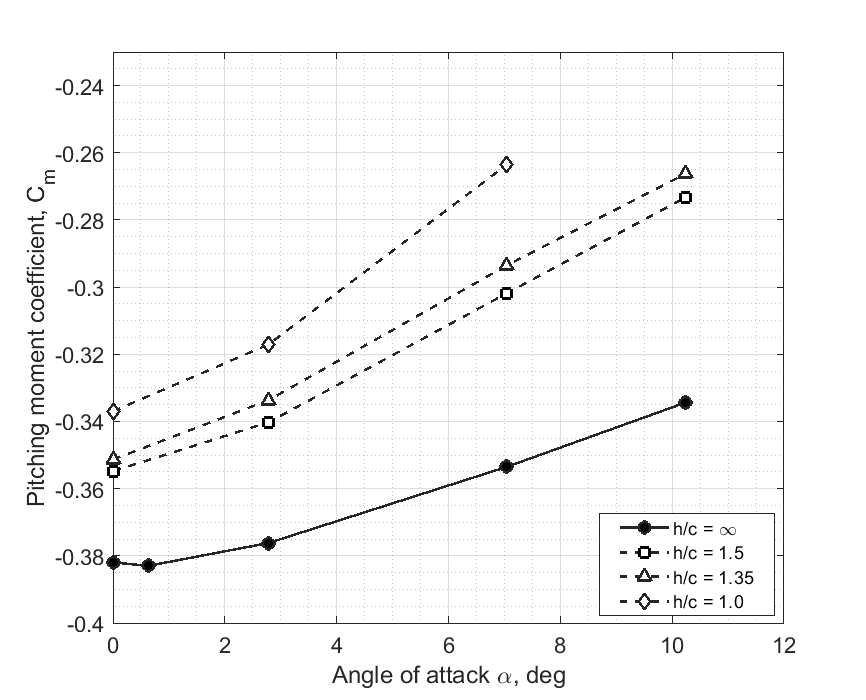}
\caption{Longitudinal aerodynamic characteristics In-Ground-Effect (IGE) at $Re=5.49\times10^6$, $M=0.2$ and flaps inboard$/$outboard$=40^\circ/37^\circ$ for $h/c = \infty, 1.5, 1.35$ and $1.0$.}
\label{Fig7-GroundEffect-pitch}
\end{figure}

The simulations are focused on analyzing the ground effect on the aerodynamic coefficients $C_L$, $C_D$, and $C_m$ at different height-to-chord ratios. The obtained computational results at different $h/c$ are shown in Fig. \ref{Fig7-GroundEffect-pitch}. At low angles of attack $\alpha < 10.0^\circ$ the maximum positive increment in the lift coefficient $\Delta C_L=0.055$ takes place at $\alpha=2.78^\circ$, which is a $4.029\%$ increment in comparison to the OGE magnitude. There are large changes in the drag and the pitching moment coefficients observed at the same closeness to the ground. As the distance to the ground reduces from $h/c=\infty$ to $h\c=1.0$, a significant decrease in the drag coefficient $C_D$ and a noticeable increase in the pitching moment coefficient $C_m$ in the nose-up direction are present. When comparing the pitching moment coefficient at $\alpha=7.045^\circ$ and $h/c=1.0$ against $h/c=\infty$ a $25\%$ increase in the nose-up pitching moment coefficient is shown. This is a quite noticeable change in the pitching moment and the drag coefficients affecting both the trimming conditions in the longitudinal motion and it's stability characteristics.

\subsection{Ground effect on the lateral-directional aerodynamics}

\begin{figure}[htb!]
\centering
\includegraphics[width=0.45\textwidth]{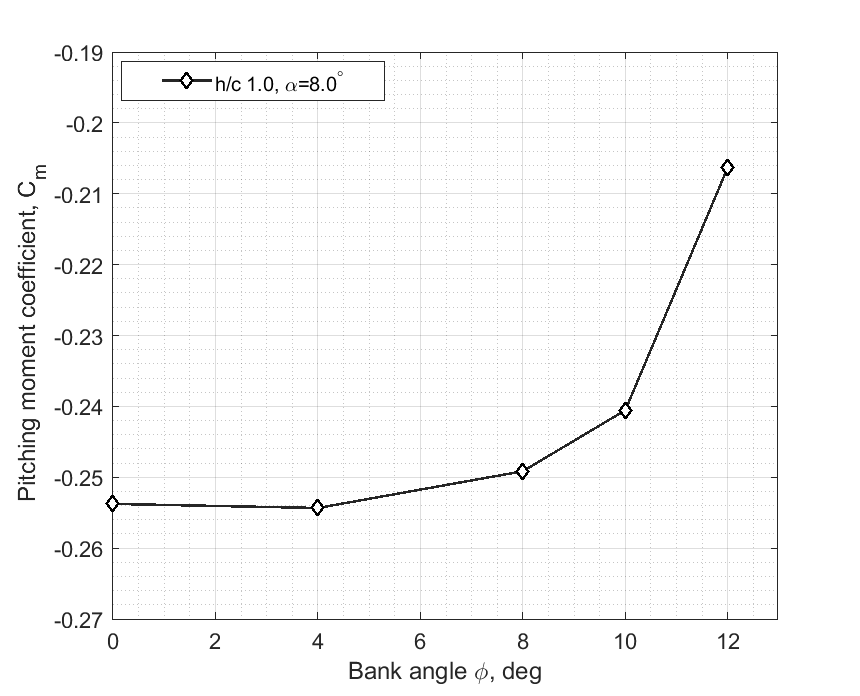}
\includegraphics[width=0.45\textwidth]{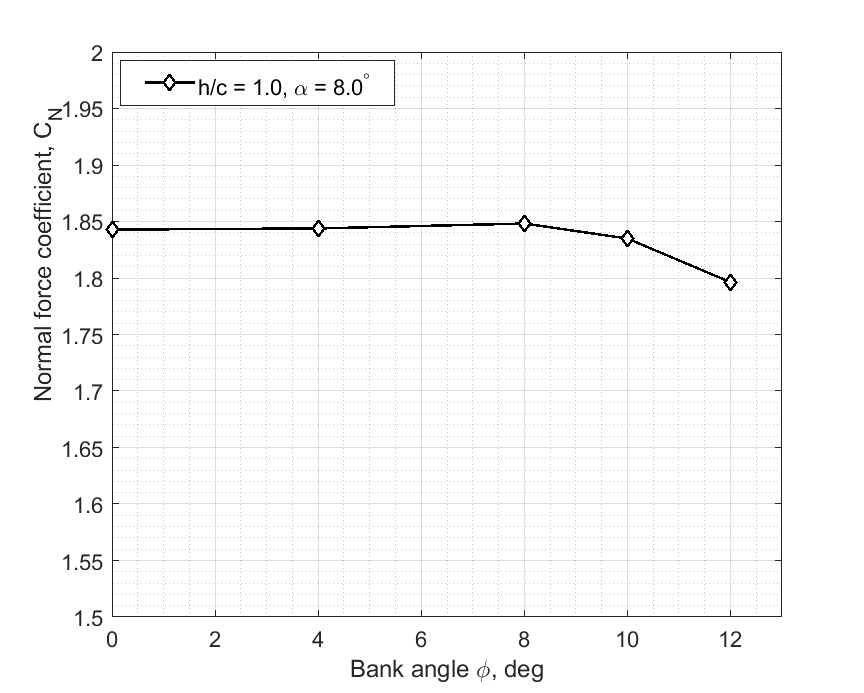}
\includegraphics[width=0.45\textwidth]{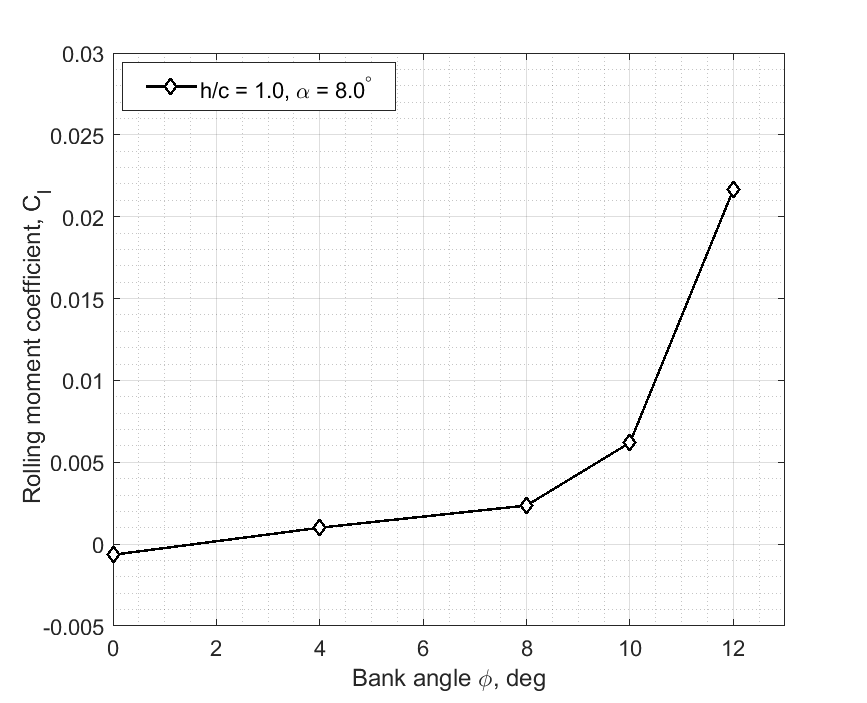}
\includegraphics[width=0.45\textwidth]{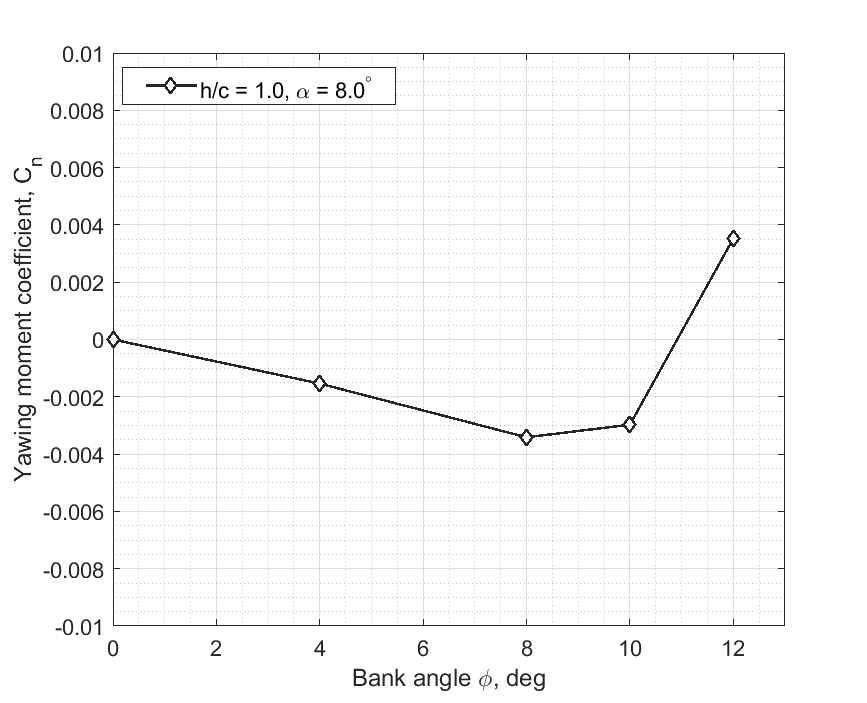}
\caption{The normal force and pitching moment coefficients along the rolling and yawing moment coefficients versus bank angle at $\alpha = 8^\circ$, $Re=5.49\times10^6$, $M=0.2$ and flaps inboard$/$outboard$=40^\circ/37^\circ$, respectively.}
\label{Fig8-GroundEffect-roll}
\end{figure}

The simulations of the ground effect on the rolling and yawing moment coefficients are conducted by introducing some non-zero bank angle. The landing under cross-wind conditions with sideslip will evoke significant bank angles due to the need to trim the aircraft \cite{FSF}. Such aircraft attitudes in close proximity to the ground will induce additional changes in the rolling and yawing moments. A better understanding of flight dynamics, trimming and stability conditions in the case of crosswind landing in close proximity to the ground is an important task from the flight safety point of view \cite{FSF}. 
The obtained simulation results for angle of attack $\alpha=8.0^\circ$ at $h/c=1.0$ and for the range of bank angles $0^\circ\leq \phi \leq 12^\circ$ are presented in Fig. \ref{Fig8-GroundEffect-roll}.

\begin{figure}[htb!]
\centering
\includegraphics[width=0.75\textwidth]{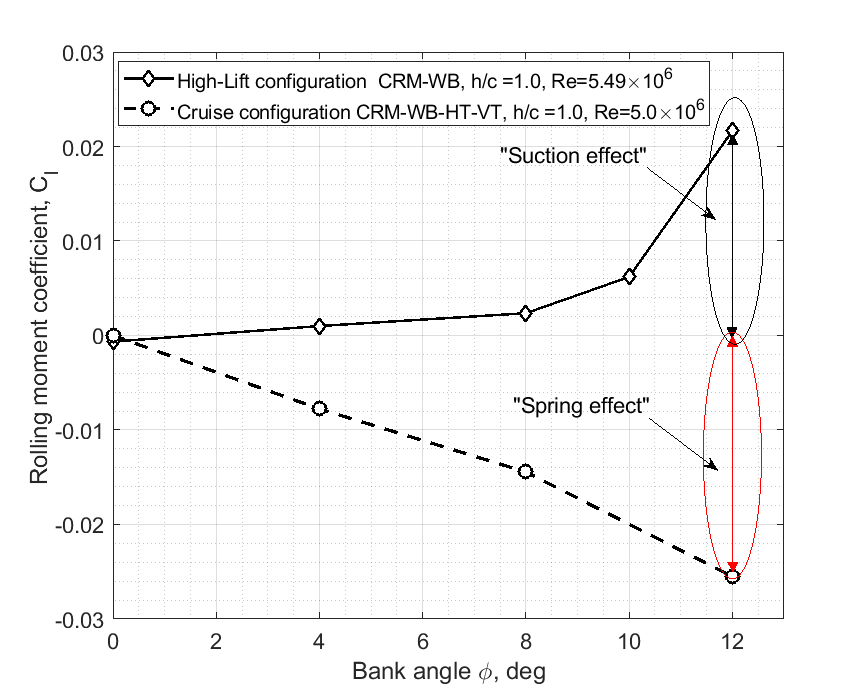}
\caption{Cruise vs high-lift CRM configurations: opposite trends in the rolling moment coefficient vs bank angle.}
\label{Fig9-CruisevsHighLiftGroundEffect-roll}
\end{figure}

The simulation results in Fig.\ref{Fig8-GroundEffect-roll} show that the pitching moment coefficient $C_m$ quite significantly increases with the increase of bank angle in close proximity to the ground (a nose-up effect). There are very small negative changes in the normal force coefficient $C_N$ at bank angles $\phi > 8.0^\circ$. For example, at $\phi=12.0^\circ$ a decrease in the normal force coefficient $C_N$ is equivalent to a loss of approximately $2.17\%$ of the normal force. Although the loss of the normal force coefficient in percentage is not significant, the loss of the normal force coefficient is throughout the right wing and increases towards the tip of the right wing. Due to the long arm on the wing tip, the small change in the normal force generates a significant contribution to the rolling moment $C_l$. This phenomenon is reflected in the trends for the rolling moment coefficient shown in Fig.\ref{Fig8-GroundEffect-roll}. With the increase of bank angle $\phi$ (right wing towards the ground) the rolling moment coefficient increases from zero at $\phi=0^\circ$ to $C_l\approx0.0218$ at $\alpha = 12^\circ$. A positive rolling moment $C_l > 0$ indicates the presence of a "suction effect" pressing the right wing to the ground.  There is a linear negative increment of the yawing moment coefficient $C_n$ up until $\phi=8.0^\circ$ followed by a leveling off at $\phi = 10.0^\circ$ and changing a sign in slope to positive at $\phi=12.0^\circ$.
 The results for the rolling moment coefficient with respect to changes in bank angle obtained for the high-lift CRM-WB configuration have the opposite trend to the cruise CRM wing-body configuration \cite{MSereezImpactofGround}. With the increase of bank angle $\phi$ the right wing rotates towards the ground and on the cruise CRM configuration, the negative rolling moments are generated causing the right wing to repel away from the ground. In this case, a "spring effect" arises. The comparison between the high-lift and the cruise CRM configurations is shown in Fig. \ref{Fig9-CruisevsHighLiftGroundEffect-roll}. 
 
\begin{figure}[htb!]
\centering
\includegraphics[width=0.98\textwidth]{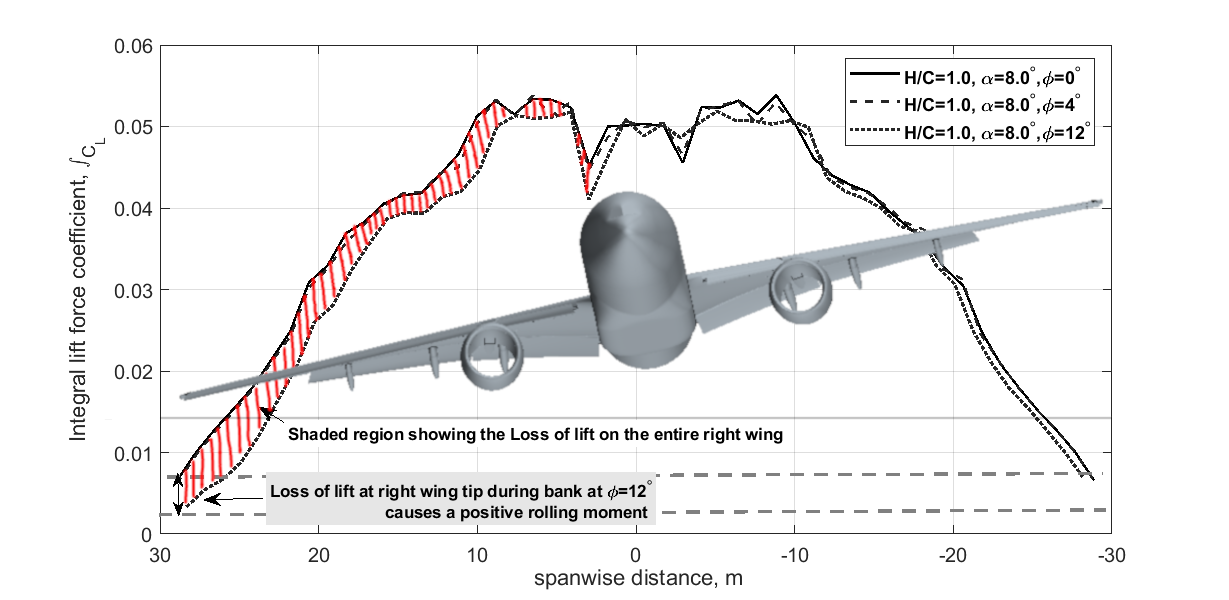}
\caption{Accumulated lift force coefficient in spanwise direction at different bank angles in close-proximity to the ground at $h/c=1.0$ and $\alpha=8.0^\circ$.}
\label{Fig10-AcLift}
\end{figure}

To illustrate the loss of the normal force coefficient throughout the span of the wing a spanwise normal force distribution is shown in Fig. \ref{Fig10-AcLift}. The solid lines demonstrate the spanwise distribution of the normal force coefficient for the $\phi=0^\circ$ case and the dashed lines represent the cases with $\phi=4^\circ$ and $\phi=10^\circ$. It is evident that the most significant loss of the normal force occurs on the right wing. And as mentioned earlier the wing tip having the longest arm generates a significant amount of the rolling moment, see Fig.\ref{Fig8-GroundEffect-roll}.

\begin{figure}[htb!]
\centering
\includegraphics[width=0.8\textwidth]{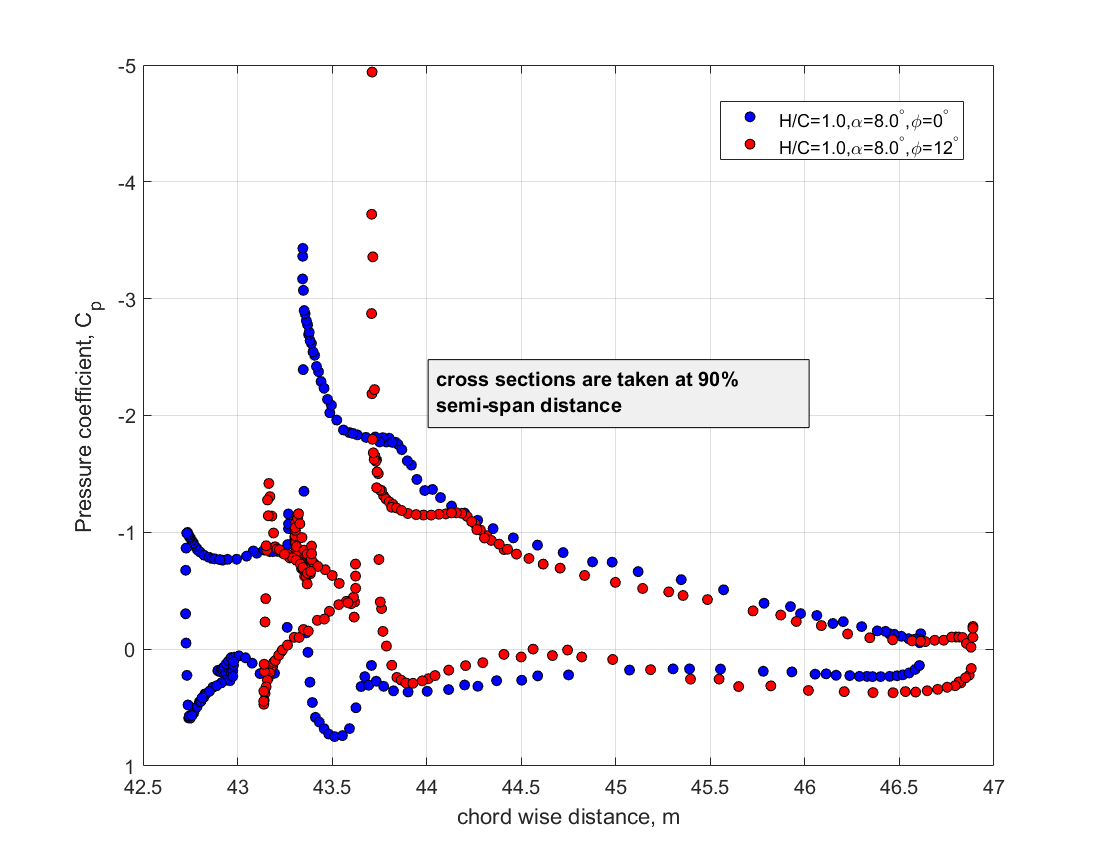}
\caption{Pressure coefficient $C_p$ distributions at $90\%$ semi-span distance for $\phi=0^\circ$ and $\phi=12.0^\circ$ in close-proximity to the ground at $h/c=1.0$ and $\alpha=8.0^\circ$.}
\label{Fig11-CpCrossSectionsETA0p9}
\end{figure}

The loss of the normal force on the wing is related to the changes in the pressure on the wing surface. In order to compare the pressure coefficient distribution at a fixed location, the coefficient  $C_p$  is plotted for a cross-section cutting the wing in the $x$ direction at $90\%$ semi-span distance. The $C_p$ plots are shown in Fig. \ref{Fig11-CpCrossSectionsETA0p9}. Blue color-filled circle markers show the $C_p$ for the case with zero bank angle and the red color-filled circles represent the $C_p$ values for the case with bank angle $\phi=12.0^\circ$. 

The most noticeable trend in this comparison is that the peak suction pressure on the main wing section is higher for the case with $\phi=12.0^\circ$ than for the case with $\phi=0^\circ$. However, this is only true for a very small portion of the leading edge of the main airfoil. It can be stated that when $\phi=12.0^\circ$, for the majority of the chordwise distance of the slats and the main wing section, lower positive pressure and also lower suction pressure are generated leading to an overall lower integral value of the normal force coefficient. 

\subsection{Post-processing of the flow field in ground effect}
\begin{figure}[htb!]
\centering
\includegraphics[width=0.72\textwidth]{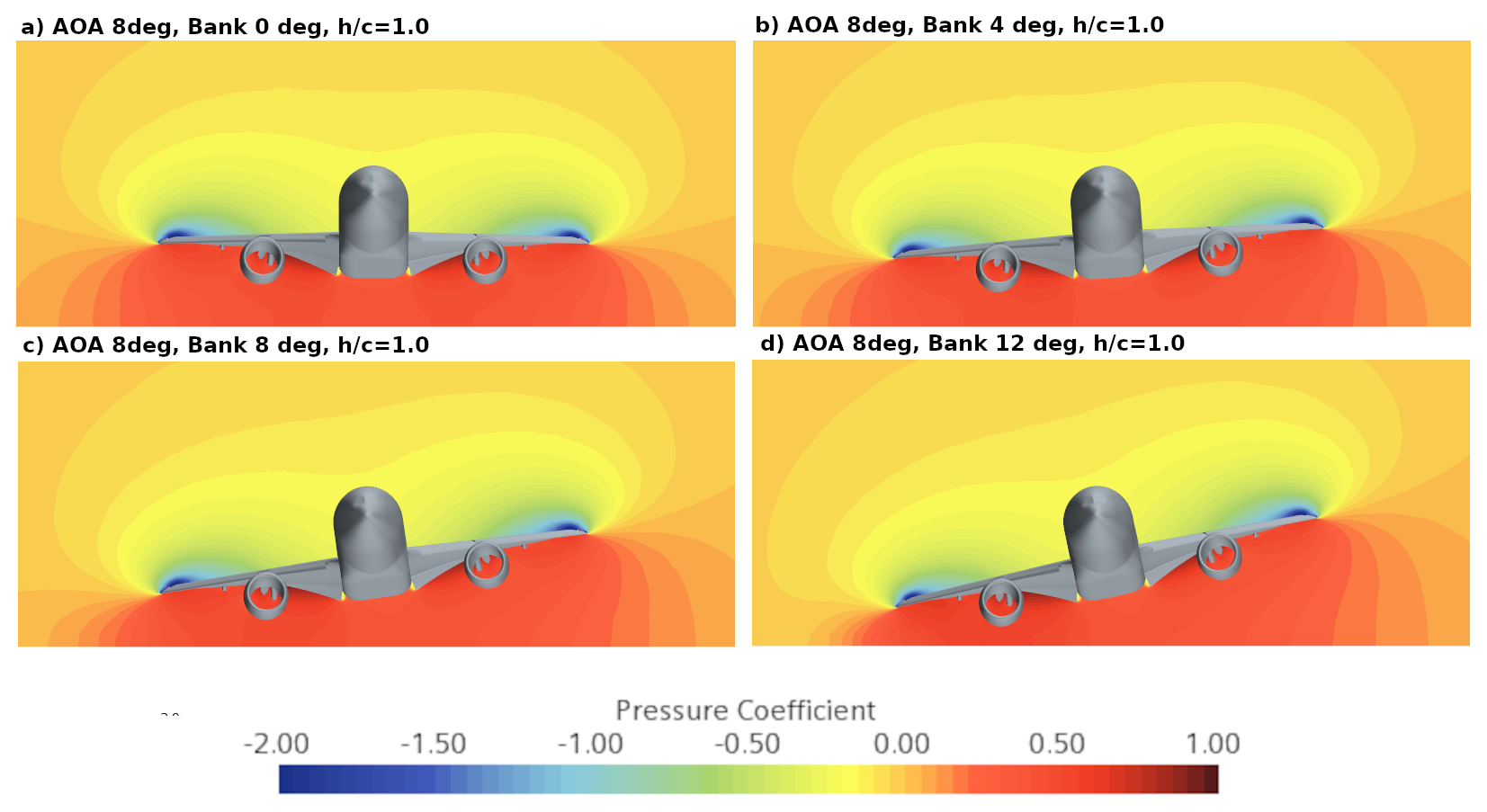}
\caption{Flow field pressure coefficient $C_p$ contours in cross-section taken at $x=37m$ in different bank configurations at $\alpha=8.0^\circ$, $Re=5.49\times10^6$,  and $M=0.2$.}
\label{Fig12-GroundEffect-visualCp}
\end{figure}

The analysis of the flow field is carried out using the Star-CCM+ CFD post-processing tools. Fig.\ref{Fig12-GroundEffect-visualCp} shows the pressure coefficient contours in a plane cross-section behind the center of gravity at $x=37m$ at a height-to-chord ratio of $h/c=1.0$ for different bank angles. When $\phi=0^\circ$, as expected, a uniform pressure coefficient distribution is observed. With the increase in bank angle, the high-pressure zone on the right wing side of the aircraft model reduces. This effect is maximum when $\phi=12.0^\circ$, leading to the loss of the normal force coefficient as seen in Fig. \ref{Fig8-GroundEffect-roll}.

\begin{figure}
       \centering
         \includegraphics[width=0.75\textwidth]{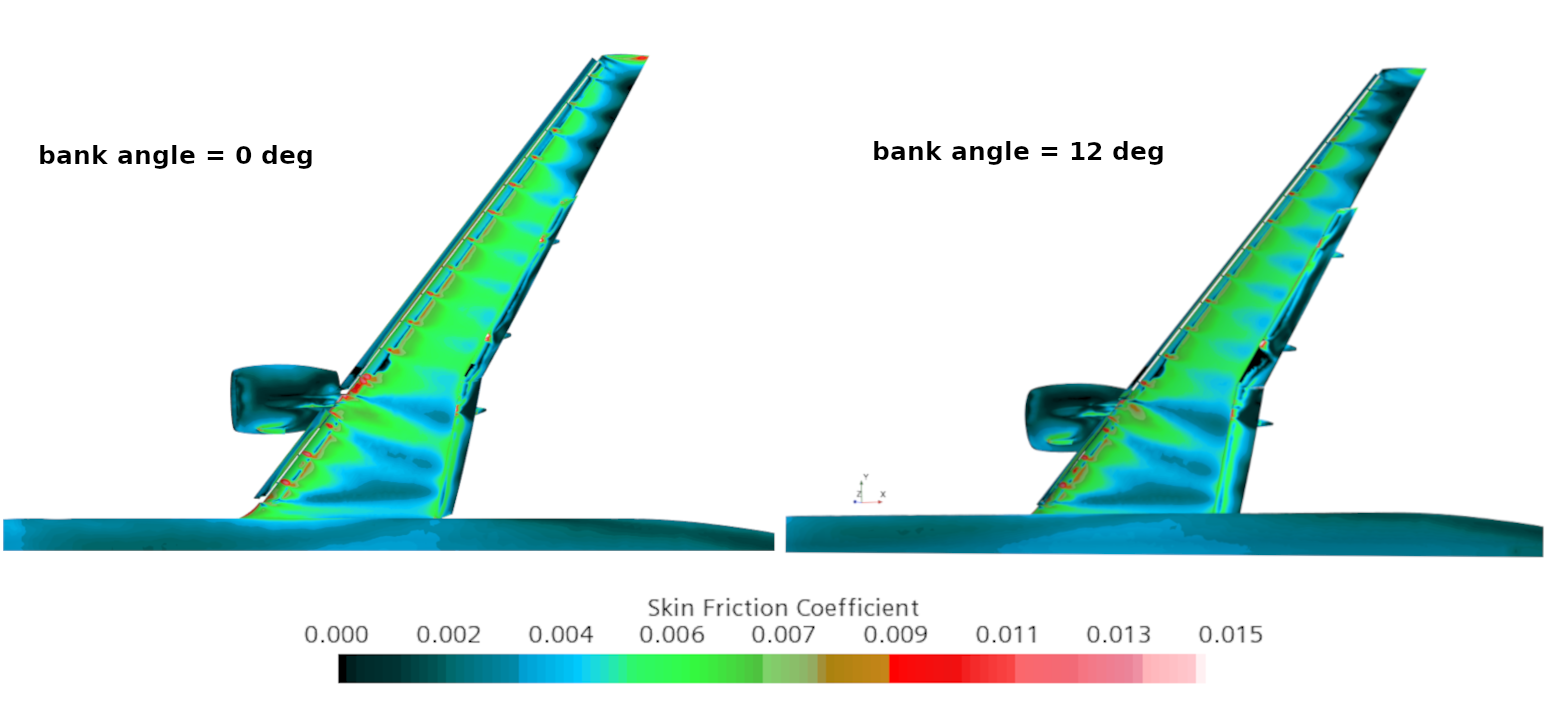}
         \caption{Skin friction coefficient $C_f$ contours on the upper surface of the wing for different bank angles at $\alpha=8.0^\circ$, $Re=5.49\times10^6$,  and $M=0.2$.}
        \label{Fig13-CFdistribution}
\end{figure}

In Fig. \ref{Fig13-CFdistribution} the skin friction coefficient $C_f$ contours on the upper surface of the wing are shown at  $h/c=1.0$ and $\alpha=8.0^\circ$. The flow separation on the upper surface of the wing is more profound when $\phi=12.0$ in comparison to the zero-bank position. More specifically there is a larger separation zone (see the black colour region in Fig. \ref{Fig13-CFdistribution} towards the tip of the wing and also on the outboard flap which also explains the loss of the normal force coefficient on the right wing seen in Fig. \ref{Fig8-GroundEffect-roll}.

\begin{figure}[htb!]
\centering
\includegraphics[width=0.7\textwidth]{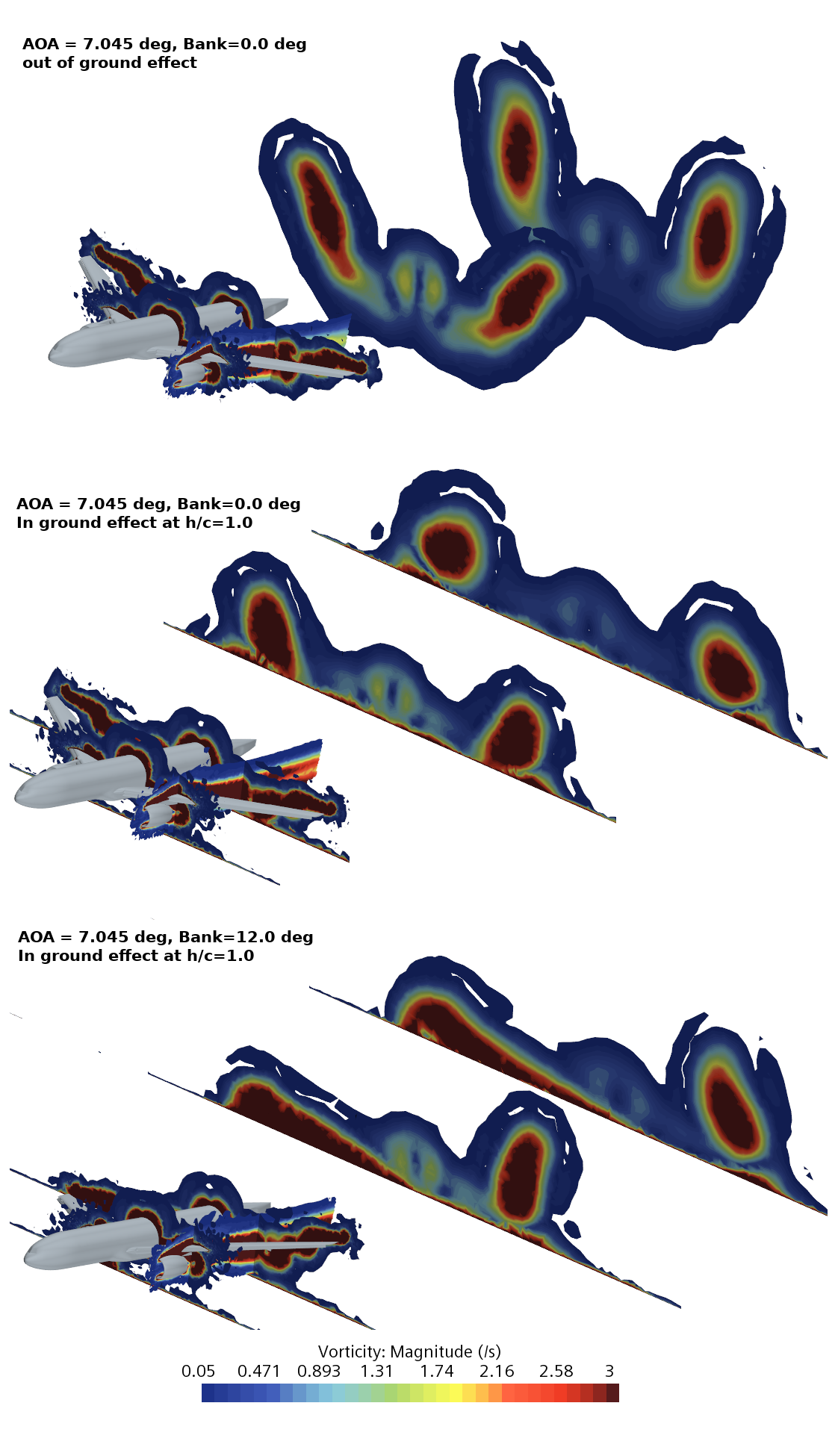}
\caption{Flow field vorticity contours behind the CRM-WB-HL configuration in out of ground and in ground effect at $\alpha=7.045^\circ$, $Re=5.49\times10^6$, $M=0.2$, flaps inboard$/$outboard$=40^\circ/37^\circ$.}
\label{Fig14-Vorticity}
\end{figure}
Further visualizations are carried out using vorticity contours projected on four distinct plane cross sections cutting through the aircraft in the x-direction and then behind the aircraft model and are shown in Fig. \ref{Fig14-Vorticity}. The changes in the vorticity and flow structure behind the wing are certainly quite interesting. When compared to the out-of-ground effect simulations, in close proximity to the ground at $h/c=1.0$ the size of the two counter-rotating vortices which are shed behind the plane decreases in their radial circumference and maintains structure but does not get distorted. The influence of downwash is quite clear in this process. At $h/c=1.0$ the vortex pair on the two sides are no longer identical and more interestingly, the vortex behind the right wing has broken down and deformed into a stretched but high-intensity vorticity zone.  At this point, it is clear that the vortex formation and their topology during the ground effect phenomenon is certainly different from that of out-of-ground effect simulations, a deeper and detailed conclusion needs further investigation. 

\section{Concluding remarks}
 The presented URANS simulations of the longitudinal and lateral-directional aerodynamics of the NASA High-Lift CRM configuration for investigation of the ground effect allow us to make the following conclusions:
\label{5-CFDConclusions}
\begin{itemize}
        \item
    The simulation results for the longitudinal characteristics in symmetric attitudes were validated against the experimental data from the QinetiQ Five-Meter Pressurized Low-Speed Wind Tunnel FL4037 without the effect of the ground and showed very good agreement at low angles of attack and the stall region.
    \item
  Ground effect simulation at non-zero bank angles has demonstrated the presence of significant changes in the rolling and yawing moment coefficients, which can significantly affect trim conditions and lateral directional stability during crosswind landings.
    \item
  Future work is planned for generating a full set of aerodynamic coefficients, including non-stationary rotational derivatives, to simulate 6-DOF crosswind landing flight using a URANS-based approach.
\end{itemize}

\bibliographystyle{unsrt}
\bibliography{citations}
\end{document}